\documentclass[prl, twocolumn,amsmath,amssymb]{revtex4}
\pdfoutput=1
\usepackage{graphicx}
\usepackage{color}
\usepackage{graphicx}
\usepackage{dcolumn}
\usepackage{bm}
\usepackage{amsmath}
\usepackage{amsfonts}
\usepackage{bbm}
\usepackage{subfigure}
\usepackage{setspace}
\newcommand{\beq}{\begin{eqnarray}}
\newcommand{\eeq}{\end{eqnarray}}

\newcommand{\bmp}{\noindent\begin{minipage}{16cm}}
\newcommand{\emp}{\end{minipage}\vskip 7mm} 

\def\drawbox#1#2{\hrule height#2pt
        \hbox{\vrule width#2pt height#1pt \kern#1pt
              \vrule width#2pt}
              \hrule height#2pt}

\def\Asym#1#2{\vcenter{\vbox{\drawbox{#1}{#2}
              \kern-#2pt 
              \drawbox{#1}{#2}}}}

\newcommand{\Om}{{\cal O}_{\widetilde{\psi}{\psi}}}
\newcommand{\dm}{ d_{\widetilde{\psi}{\psi}}}

\begin{document}
\title{\Large  Conformal Chiral Dynamics}
\author{Francesco {\sc Sannino}}
\email{sannino@ifk.sdu.dk}
\affiliation{High Energy Center, University of Southern Denmark, Campusvej 55, DK-5230 Odense M, Denmark.}


\begin{abstract}
We investigate the chiral dynamics of gauge theories developing an infrared stable fixed point.  We determine the dependence of the bilinear fermion condensate on the underlying fermion mass and its anomalous dimension.  We introduce the instanton contributions and investigate how they affect the dynamics near the fixed point. We generalize the Gell-Mann Oakes Renner relation and suggest to use  it to {\it uncover} the presence of an infrared fixed point of the underlying gauge theory. Our results have an immediate impact on the construction of sensible extensions of the Standard Model of particle interactions and the general understanding of the phase diagram of strongly coupled theories.
 \end{abstract}

\maketitle
Non-abelian gauge theories exist in a number of  distinct phases which can be classified according to the characteristic dependence of
the potential energy on the distance between 
two well separated static sources.  The collection of all of these different behaviors,
when represented, for example, in the  flavor-color space, constitutes the {\it Phase Diagram} of the given gauge theory. 

Knowing the phase diagram of strongly coupled theories has an immediate impact on the construction of sensible extensions of the Standard Model (SM) of particle interactions \cite{Sannino:2008ha}. Dynamical breaking of the electroweak symmetry is a time-honored example \cite{Weinberg:1979bn,Susskind:1978ms}.  The use of fermions transforming according to higher dimensional representations of the
new gauge group is leading to several interesting phenomenological possibilities \cite{Sannino:2004qp,Dietrich:2005jn,Dietrich:2006cm} such as Minimal Walking Technicolor (MWT)\cite{Foadi:2007ue} and Ultra Minimal Walking Technicolor (UMT) \cite{Ryttov:2008xe}. 
 These models lie close, in theory space, to theories with non-trivial infrared fixed points (IRFP)s \cite{Sannino:2004qp,Ryttov:2007cx}.  In the vicinity of such a zero of the beta-function the coupling constant flows slowly, i.e.  {\it walks} \cite{Eichten:1979ah,Holdom:1981rm,Yamawaki:1985zg,Appelquist:1986an}. Knowledge of the phase diagram is relevant also to provide natural ultraviolet completions of unparticle \cite{Georgi:2007ek} models \cite{Sannino:2008nv,Sannino:2008ha}. To gain analytic insight one can use the conjectured 
all-order beta function for nonsupersymmetric gauge 
theories \cite{Ryttov:2007cx} together with the 
constraints from the unitarity of the conformal operators \cite{Mack:1975je}.
Other approaches  are based on the older truncated Schwinger-Dyson
approach (SD) \cite{Appelquist:1988yc} or more recent ideas \cite{Alkofer:2004it}.
The analytical phase diagram obtained by this approach,
and a comparison of it to recent lattice results \cite{Catterall:2007yx,Catterall:2008qk,
Shamir:2008pb,DelDebbio:2008wb,DelDebbio:2008zf, Hietanen:2008vc,Appelquist:2007hu,Deuzeman:2008sc,Fodor:2008hn},
is summarized in \cite{Sannino:2008ha}. 

The goal here is to study the chiral properties at the IRFP, i.e. the {\it conformal chiral dynamics}. 

\begin{figure}[h]
\begin{center}
\includegraphics[width=6cm,]{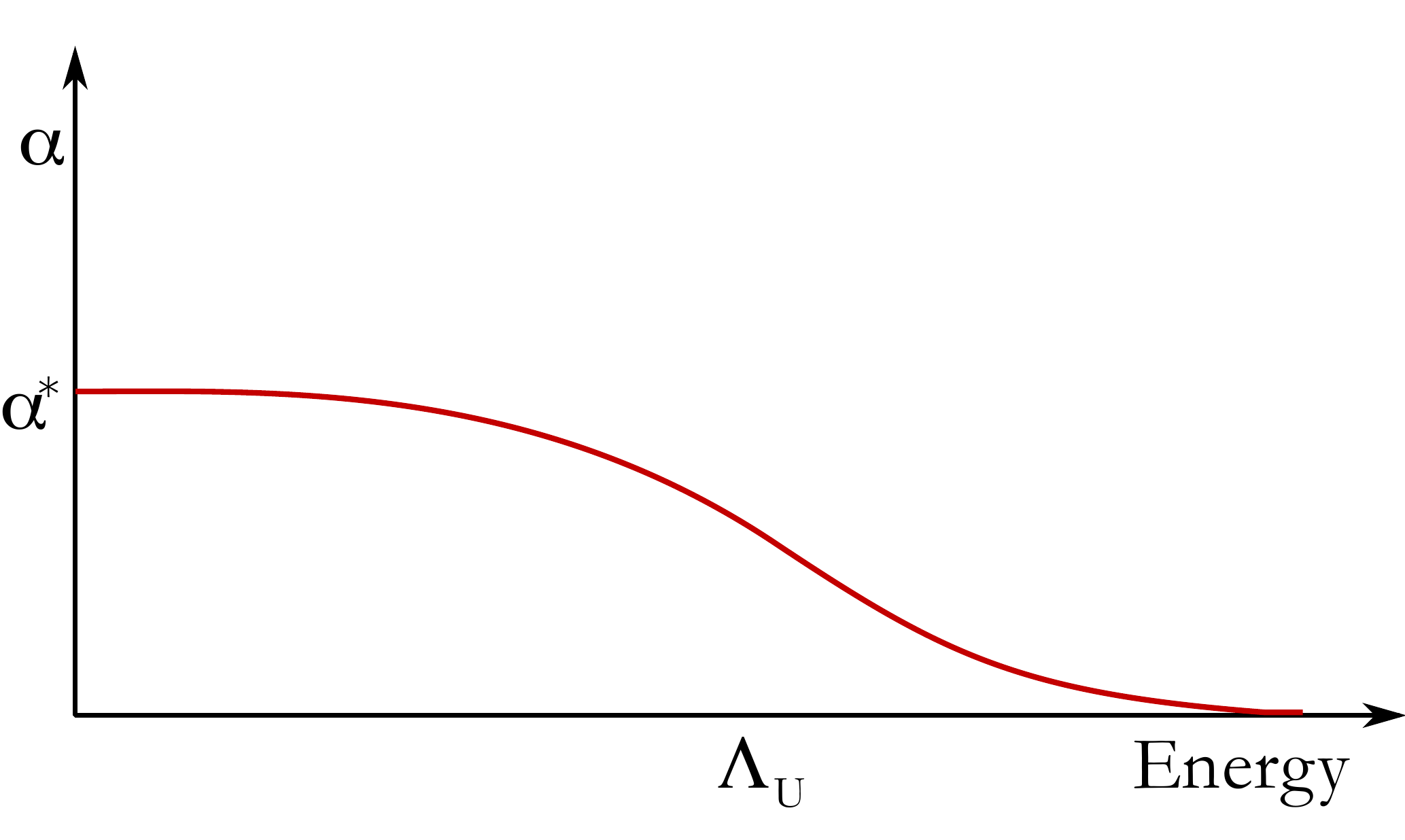}
\caption{ Running of the coupling constant in an asymptotically free gauge theory developing an infrared fixed point for a value $\alpha = \alpha^{\ast}$. }
\label{fig1}
\end{center}
\end{figure}

\vskip .1cm
\noindent
{\bf  Conformal Chiral Dynamics:}
Our starting point is a nonsupersymmetric non-abelian gauge theory with sufficient massless fermionic matter to develop a nontrivial IRFP. The cartoon of the running of the coupling constant is represented in Fig.~\ref{fig1}. In the plot $\Lambda_U$ is the dynamical scale below which the IRFP is essentially reached. It can be defined as the scale for which $\alpha$ is $2/3$ of the fixed point value in a given renormalization scheme. 
If the theory possesses an IRFP the chiral condensate must vanish at large distances. Here we want to study the behavior of the condensate when a flavor singlet mass term is added to the underlying Lagrangian: 
\begin{eqnarray}
\Delta L = - m\,{\widetilde{\psi}}{\psi} + {\rm h.c.} \ ,
\end{eqnarray} 
with $m$ the fermion mass and $\psi^f_{c}$ as well as $\widetilde{\psi}_f^c$ left transforming two component spinors, $c$ and $f$ represent color and flavor indices.  The omitted color  and flavor indices, in the Lagrangian term, are contracted. 
We consider first the case of fermionic matter in the fundamental representation of the $SU(N)$ gauge group. We then generalize our results to the case of higher dimensional representations.  

The effect of such a term is to break the conformal symmetry together with some of the global symmetries of the underlying gauge theory.  The composite operator:
\begin{equation}
{{\cal O}_{\widetilde{\psi}{\psi}}}^{f^{\prime}}_f  = \widetilde{\psi}^{f^{\prime}}{\psi}_f  \ , 
\end{equation}
has mass dimension $\displaystyle{
d_{\widetilde{\psi}{\psi}} = 3 - \gamma_m}$ with 
$\gamma_m$  the anomalous dimension of the mass term. At the fixed point $\gamma_m$ is a positive number smaller than two \cite{Mack:1975je}. We assume $m \ll \Lambda_U$.  Dimensional analysis demands:
\begin{eqnarray}\Delta L  \rightarrow 
-m\, \Lambda_U^{\gamma_m} \, {\rm Tr}[\Om] + {\rm h.c.} \ .
\label{mass}
\end{eqnarray}
The mass term is a relevant perturbation around the IRFP driving the theory away from the fixed point.  It will induce a nonzero vacuum expectation value for $\Om$ itself proportional to $\delta^{f^\prime}_f$. It is convenient to define ${\rm Tr}[\Om]  = N_f  \cal O $ with $\cal O$ a flavor singlet operator.  The relevant low energy Lagrangian term is then: 
\begin{equation}
-m\, \Lambda_U^{\gamma_m} \, N_f \cal O + {\rm h.c.}  \ .
\end{equation}
To determine the vacuum expectation value of $\cal{O}$ we replace it, formally, with a sum over an infinite number of canonically normalized single particle states \cite{Stephanov:2007ry}:
\begin{equation}
{\cal O}(x) = \sum_{n=1}^{\infty} f_n \varphi_n (x) \ .
\end{equation}
Each state possesses a mass $M_n$ whose value is controlled by an artificial mass gap $\Delta$ and a function of $n$, call it $z(n)$, with the properties $z(n+1)>z(n)$ and $z(1)=1$. 
\begin{equation}
M^2_n = \Delta^2 z(n) \ .
\end{equation}
We also have \cite{Stephanov:2007ry}: 
\begin{equation}
f_n^2 = {\cal F}_{\dm} \frac{dz(n)}{dn}\, \Delta^2 (M^2_n)^{\dm -2} \ ,
\end{equation}
with $ {\cal F}_{\dm}$ a function depending on the scaling dimension of the operator as well as the details of the underlying dynamics \cite{Sannino:2008nv}.  
Because of the presence of the fictitious mass terms the potential reads:
\begin{eqnarray}
V &=& m\, \Lambda_U^{\gamma_m} \, N_f \sum_{n=1}^{\infty} f_n \varphi_n   +  \bar{m}\, \Lambda_U^{\gamma_m} \, N_f \sum_{n=1}^{\infty} f_n \bar{\varphi}_n  \nonumber \\
&+& \sum_{n=1}^{\infty} M^2_n \varphi_n \bar{\varphi}_n  \ .
\end{eqnarray}
The bar over the fields and the fermion mass indicates complex conjugation. The extremum condition yields:
 \begin{equation}
 \langle  \bar{\varphi}_n\rangle = - m \Lambda_U^{\gamma_m} N_f \frac{f_n}{M^2_n} \ ,
 \end{equation}
yielding: 
\begin{equation}
\langle {\cal O} \rangle =  - \bar{m} \Lambda_U^{\gamma_m} N_f \sum_{n=1}^{\infty}  \frac{f_n^2}{M^2_n}  \ .
\end{equation}
We now take the limit $\Delta^2 \rightarrow 0$ and the sum becomes an integral. {}For any specific  function $z(n)$ it is easy to show that:
\begin{equation}
\langle {\cal O} \rangle =  - \bar{m} \Lambda_U^{\gamma_m} N_f {\cal F}_{\dm}\Omega\left[\Lambda_{UV} ,\Lambda_{IR}\right] \ , 
\label{Ovev}
\end{equation}
 with 
 \begin{equation}
 \Omega\left[\Lambda_{UV} ,\Lambda_{IR}\right]  = \frac{1}{1-\gamma_m} \left[ \Lambda_{UV}^{2(1-\gamma_m)} - \Lambda_{IR}^{2(1-\gamma_m)} \right] \ .
 \end{equation}
 The  ultraviolet and infrared cutoffs are introduced to tame the integral in the respective regions. A simple physical interpretation of these cutoffs is the following. At very high energies, at scales above $\Lambda_{U}$, the underlying theory flows to the ultraviolet fixed point and we have to abandon the description in terms of the composite operator. This immediately suggests that $\Lambda_{UV}  $ is naturally identified with $\Lambda_U$. The presence of the mass term induces a mass gap, which is the quantity we are trying to determine. The induced physical mass gap is a natural infrared cutoff. We, hence, identify  $\Lambda_{IR} $ with the physical value of the condensate. We find:
 \begin{eqnarray}
 \langle \widetilde{\psi}^f_c \psi^c_f \rangle  &\propto& -m \Lambda_U^2 \ ,  \qquad ~~~~~~0 <\gamma_m  < 1 \ , \label{BZm} \\
 \langle \widetilde{\psi}^f_c \psi^c_f \rangle  &\propto &   -m \Lambda_U^2  \log \frac{\Lambda^2_U}{|\langle {\cal O} \rangle|}\ , ~~~   \gamma_m \rightarrow  1    \ , \label{SDm} \\
 \langle \widetilde{\psi}^f_c\psi^c_f \rangle   &\propto &  -m^{\frac{3-\gamma_m} {1+\gamma_m}} 
 \Lambda_U^{\frac{4\gamma_m} {1+\gamma_m}}\ , ~~~1<\gamma_m  \leq 2 \ .
   \label{UBm}
 \end{eqnarray}
We used  $\langle \widetilde{\psi} \psi \rangle \sim \Lambda_U^{\gamma_m} \langle {\cal O} \rangle $ to relate the expectation value of ${\cal O}$ to the one of the fermion condensate. Via an allowed axial rotation $m$ is now real and positive. 
It is instructive to compare these results with the ones obtained via naive dimensional analysis (NDA)  \cite{Fox:2007sy} also discussed in \cite{Luty:2008vs} and in \cite{Sannino:2008nv}. We find:
\begin{equation}
 \langle \widetilde{\psi}^f_c \psi^c_f \rangle _{\rm NDA} \propto  -m^{\frac{3-\gamma_m} {1+\gamma_m}} 
 \Lambda_U^{\frac{4\gamma_m} {1+\gamma_m}} \ . 
\end{equation}
Note that one recovers the previous scaling as function of $m$ (up to logarithmic corrections) only for $1\le \gamma_m \le 2$. The failure of NDA for a smaller anomalous dimension is due to the fact that  the ultraviolet physics is not captured by NDA \cite{Sannino:2008nv}. 

\noindent
{\bf  Instantons:}
The underlying gauge theory suffers of an axial anomaly resulting in an explicit breaking of the $U_A(1)$ symmetry.  {}To take into account this phenomenon we add the following instanton-induced term to the potential of the theory \cite{Witten:1979vv,Veneziano:1979ec,Witten:1980sp,Rosenzweig:1979ay,DiVecchia:1980ve,Nath:1979ik,Kawarabayashi:1980dp,Shifman:1986zi,'tHooft:1986nc,Sannino:1999qe}:
\begin{equation}
c\,\Lambda_U^{4}\,\frac{{\rm det} \left[\Om\right]}{\Lambda_U^{\dm\, N_f}} +{\rm h.c.} = c\,\frac{ {\cal O}^{N_f}}{\Lambda_U^{(\dm\, N_f - 4)}} +{\rm h.c.} \ .
\end{equation}
In terms of the fields $\varphi_m$ the VEV equation reads: 
\begin{equation} 
\langle \varphi_\ell \rangle = - \frac{f_\ell }{M^2_\ell} \left[\bar{m} N_f \Lambda_U^{\gamma_m} + \bar{c} \, N_f  \frac{\left(\sum_n f_n \langle \bar{\varphi}_n \rangle\right)^{N_f -1}}{\Lambda_U^{\dm \,N_f -4}}\right] \ .
\end{equation}
We search for a solution of the previous equation of the form
$\displaystyle{\langle \varphi_\ell \rangle =  a \, \frac{f_\ell}{ M^2_\ell}} $.
Substituting in the previous expression we deduce:
\begin{equation}
\bar{a}^{N_f -1} \bar{C} + a + \bar{M} = 0 \ , 
\label{master}
\end{equation}
with 
\begin{eqnarray}
\bar{C} & = & 
 \frac{ \bar{c} N_f}{\Lambda_U^{\dm \,N_f -4}} \left( {\cal F}_{\dm}\Omega\left[\Lambda_{U} ,\Lambda_{IR}\right]  \right)^{N_f -1}\ , \label{Ccoef} \\
 \bar{M} & =  &\bar{m} N_f \Lambda_U^{\gamma_m} \ .
 \label{Mcoef}
 \end{eqnarray}
 We have already taken the $\Delta\rightarrow 0$ limit in \eqref{Ccoef}. We analytically solve for $a$ in the two extreme cases, i.e., no instanton contribution ($c=0$): 
\begin{eqnarray}
{|a|} & = &|M| \ ,  \quad  \delta_a   = \pi(1+2k) - \delta_M  ~{\rm with}~ c = 0 \ . 
 \end{eqnarray}
$\delta_a$ and $\delta_M$ are the phases respectively of $a$ and $M$. This solution was found above. In the limit in which the instanton term dominates over the linear term in $a$ the solution is:
\begin{eqnarray}
\frac{|a|}{|M|} = \left( \frac{|M|^{2-N_f}}{|C|} \right)^{\frac{1}{N_f - 1}},  \quad \delta_a = \frac{\delta_M - \delta_c -\pi(1 - 2k)}{N_f -1 } \ . \nonumber \\
\end{eqnarray}
$k$ is an integer. In the instanton dominated (ID) limit:
\begin{equation}
|\langle {\cal O} \rangle_{\rm ID}| =\left( {\frac{|M|\,\Lambda_u^{\dm N_f -4}}{|c|\,N_f}}\right)^{\frac{1}{N_f - 1}} \ .
\end{equation}
The $ {\cal F}_{\dm}\Omega\left[\Lambda_{U} ,\Lambda_{IR}\right]$ term cancels. Explicitating the mass term dependence: 
\begin{equation}
 \langle \widetilde{\psi}^f_c \psi^c_f \rangle _{\rm ID} \propto \left(\frac{m}{\Lambda_U}\right)^{\frac{1}{N_f - 1}} \Lambda_U^{3} \ .
 \end{equation}
This ID contribution dominates for large values of the fermion mass . 

{}In the small mass regime we can solve \eqref{master} perturbatively in the mass. This expansion is well defined for $ 0<\gamma_m \leq 1$ since here the  $C$ coefficient is not affected by the IR divergence.  To the next leading term in $m$: 
\begin{equation}
\langle {\cal O} \rangle \simeq  -(\bar{M}  +  (-M)^{{N_f - 1}}\, \bar{C}) \ {\cal F}_{\dm}\Omega\left[\Lambda_{UV} ,\Lambda_{IR}\right] \ .
\label{Ovev2}
\end{equation}

The introduction of $\theta$-term
is a source of strong $CP$ violation appearing at low energies via the identification
$\displaystyle{\delta_M = \delta_m + \omega}$ and $ \displaystyle{ \delta_c= \theta + N_f \omega}$
with $\delta_m$ the phase of the fermion mass and $\omega$ an overall axial rotation.  Using $\omega$ one can rotate away one of the two phases but not both simultaneously. The net result is the presence in the action of ${\theta}_{eff} = \theta - N_f \delta_m$ and we find $\displaystyle{\langle {\cal O} \rangle_{\rm ID}= {\rm Exp}[-i \frac{\theta_{eff} -\pi(1-2k) }{N_f -1 }]|\langle {\cal O} \rangle_{\rm ID}|}$. According to the all-order beta function \cite{Ryttov:2007cx} an anomalous dimension smaller than one requires, for fermions in the fundamental representation,  a number of flavors larger than $11N/3$. Even for $N=2$ the number of flavors needed is sufficiently large to predict that the instanton corrections are negligible.  In the case of  fermionic matter transforming according to higher dimensional representations  the effects of the instantons are more relevant since fewer flavors are needed to reach the conformal window \cite{Sannino:2004qp} reducing the exponent of the instanton induced term. Besides, for any representation, in the region $1<\gamma_m \leq 2$, instantons are relevant as we shall demonstrate below. 
 
 \vskip .1cm
 \noindent
 { \bf 2 Dirac Fermions, 2-index Symmetric Representation of SU(3):}
There are lattice indications that this theory may develop an IRFP \cite{Shamir:2008pb}. The instanton induced potential term here has the lowest possible exponent, i.e. $N_f=2$. The all-order beta function predicts $\gamma_m=1.3$. This is the regime where the instanton term cannot be neglected. We obtain: 
 \begin{equation}
\langle {\cal O} \rangle = \frac{2m \,e^{i\delta_a}\Lambda_U^{\gamma_m}  {\cal F}_{\dm}\Omega\left[\Lambda_{UV} ,\Lambda_{IR}\right]}{\cos \delta_a + \frac{2|c|\cos (\delta_a + \delta_c)}{\Lambda_U^{2\dm - 4}} {\cal F}_{\dm}\Omega\left[\Lambda_{UV} ,\Lambda_{IR}\right]}\ , 
\label{Ovevsu2}
\end{equation}
where $m>0$, $\theta_{eff}=0$,  $\delta_M = \pi$ and $\sin \delta_a = |C|\sin(\delta_a + \delta_c)$ determines $\delta_a$. {}For $\gamma_m > 1$  $\Omega$ is IR dominated leading to:
\begin{eqnarray}
 \langle \widetilde{\psi}^f_{\{c_1,c_2\}} \psi^{\{c_1,c_2\}}_f \rangle  = -\frac{m}{c} \Lambda_U^2 + O(m^2) \  ,  \end{eqnarray}
 rather than
 \begin{eqnarray}
 \langle \widetilde{\psi}^f_{\{c_1,c_2\}} \psi^{\{c_1,c_2\}}_f \rangle  
 \propto - \left(\frac{m}{\Lambda_U} \right)^{\frac{3-\gamma_m}{1+\gamma_m}}\Lambda_U^3  \ ,  \end{eqnarray}
valid for $1<\gamma_m \leq 2$ but without the instanton term. Without using the information of the beta function one might still imagine the possibility that the anomalous dimension is smaller than one. In this case $\Omega$ is UV divergent, we replace $\Lambda_{UV}$ with $\Lambda_U$ and find that the dependence of the condensate on the mass is still linear.

\vskip .1cm
\noindent
{\bf2 Dirac Fermions in the Adjoint Representation of SU(2):}
This theory is also being investigated on the lattice \cite{Catterall:2007yx,Catterall:2008qk,DelDebbio:2008wb,DelDebbio:2008zf,Hietanen:2008vc}. Here the instanton induced term has $2N_f  = 4$ as exponent and one can still solve analytically for the condensate. If the underlying conformal theory possesses an IRFP according to the all-order beta function $\gamma_m = 3/4$ then the condensate has a linear dependence on the fermion mass. It has a dependence proportional to $m^{\frac{1}{3}}$  for larger values of the mass. 

\vskip .1cm
\noindent
{\bf Conformal Pions:}
At any nonzero value of the fermion mass the chiral and conformal symmetries are explicitly broken and single particle states emerge at low energies. The relevant ones here are the {\it conformal pions}, i.e. the would be Goldstones which in the limit of zero fermion mass cannot be described  via single particle states. We identify them via
\begin{eqnarray}
\langle {\Om}^{f^\prime}_{f}\rangle &=&  \langle {\cal O} \rangle \, U \quad {\rm with} \quad U=e^{i \frac{\pi}{F_{\pi}}}  \ . 
\label{U}
\end{eqnarray}
$\pi=\pi^a T^a$ and $T^a$ are the set of broken generators normalized according to ${\rm Tr} \left[T^a T^b\right] = \delta^{ab}1/2$. Substituting \eqref{U} in \eqref{mass}  and expanding up to the second order in the pion fields we have: 
\begin{eqnarray}
m^2_{\pi} F_\pi^2 =  -m\,\Lambda_U^{\gamma_m}\langle {\cal O}[m]\rangle \ .
\end{eqnarray}
Having determined the dependence on $m$ of $\langle O [m]\rangle$ the above generalizes the similar one in QCD \cite{Glashow:1967rx,GellMann:1968rz,Dashen:1969eg} known as the Gell-Mann Oakes Renner (GMOR)  relation. This relation can be used to {\it discover} and classify, in a physical way, different conformal fixed points. {}For example for the theories investigated above  we expect, for a very small fermion mass, $m^2_{\pi} F_\pi^2= m^2 \Lambda_U^2$. At larger masses the scaling is different  for the three cases and it can be easily deduced from our results. A similar effective Lagrangian was introduced in \cite{Sannino:2008nv}.  Assume now that the underlying gauge theory has not developed an IRFP. In this case there are only two possibilities: i) chiral symmetry breaks spontaneously yielding a condensate whose leading term in $m$ is a constant; ii) chiral symmetry is intact but a scale is still generated. Chirally paired partners emerge together with massless composite fermions appearing to saturate the 't Hooft anomaly matching conditions.  One can investigate the finite volume effects using the conformal pion lagrangian in the $\epsilon$-regime \cite{Gasser:1987ah}.

We presented a novel analysis of the nonperturbative physics related to the chiral dynamics of theories possessing an IRFP. The results are testable via first principles lattice simulations. Deviations from the QCD-like GMOR relation can be used to disentangle conformal dynamics from non-conformal dynamics. The low energy effective theories presented here, for the  {\it conformal pions},  can be extended to describe dynamical breaking of the electroweak symmetry featuring nearly conformal dynamics such as MWT and UMT. The signals from MWT at the LHC are being investigated while UMT also leads to interesting dark matter candidates \cite{Ryttov:2008xe}.

%

\end{document}